\documentclass[prl,aps,twocolumn]{revtex4}
\usepackage{epsfig}
\usepackage{appendix}

\begin{document}

\title{Statistical Mechanics of the Anomalous Behavior of Tetrahedral Liquids}
\author{Itamar Procaccia$^1$ and Ido Regev$^{1,2}$}
\affiliation{$^1$ Department of Chemical Physics, The Weizmann Institute of Science, Rehovot 76100, Israel\\
$^2$ Theoretical Division and CNLS, Los Alamos National Laboratory, Los Alamos 87545, New Mexico}
\begin{abstract}
Tetrahedral liquids such as water and silica-melt show unusual thermodynamic behavior such as a density maximum and an increase in specific-heat when cooled to low temperatures. There is a debate in the literature whether these phenomena stem from a phase transition into a low-density and high-density liquid phases, which occur in the supercooled regime. Here we consider a model of tetrahedral liquids for which we construct a volume-constrained statistical mechanical theory which quantifies the local structure of the liquid. We compare the theory to molecular dynamics simulations and show that the theory can rationalize the simulations semi-quantitatively.
We show that the anomalous density and specific heat behavior arise naturally from this theory without exhibiting a liquid-liquid phase-transition. We explain that this theory may or may not have a phase transition, depending on the volume and temperature dependence of the energy and entropy which are sensitive to small changes in the parameters of the model.

\end{abstract}

\date{\today}

\maketitle

``Tetrahedral liquids" are composed of molecules which tend to form strong bonds with four neighbors in a tetrahedral geometry. Several materials share this type of interaction, most notably water, silicon and silica. These liquids exhibit non-trivial thermodynamics at low temperature: when cooled, instead of steadily reducing in volume, they have a density maximum; the specific heat, usually a result of thermal fluctuations, increases rather than decreases and the thermal expansion becomes smaller and even reaches negative values. There is a debate in the literature whether these phenomena stem from a first-order phase transition into a low-density and high-density liquid phases or a continuous change \cite{98MS}. Extensive studies used experiments \cite{10BSSWF,98MS}, simulations \cite{03SA,09GW} and theory to clarify this point. Several groups have invoked different model descriptions which incorporate Hamiltonian terms that were believed to represent important physical effects, cf. Refs \cite{10SMSF,98T} just to mention a few. A more comprehensive approach was used by Truskett at al. \cite{02TD} which used a statistical mechanical model to describe the properties of the two-dimensional Mercedez-Benz model. Despite of these efforts the origin of the anomalies is not completely clear.

Here we apply statistical mechanics methods to study the issue on the basis of the three-dimentional Stillinger-Weber model Hamiltonian, which is a generic, empirical model of tetrahedral liquids. The Stillinger-Weber model incorporates two and three-body terms; it was successfully employed to study the properties of silicon \cite{84SW,96FA} (just to name a few), and has recently been used as a coarse-grained model of water \cite{09MM}. In this letter, we examine this system and identify the different configurations that can occur around a particle (its first coordination shell). We then assign energies, volumes and degeneracies to each possible configuration, or ``species'', using simple theoretical considerations. Finally, we calculate the free-energy of an ensemble of these species under constant volume and temperature. Since our calculation assumes constant volume and not constant pressure, it is natural to consider experiments were the liquid is confined to nano-pores \cite{07MBCFMVLMC}. We demonstrate that such a theory, based on a statistical mechanical description of the {\em local structure} of the liquid can explain the anomalous behavior of tetrahedral liquids due to a competition between energy and entropy. We also explain why the same mechanism that is responsible for the anomalous behavior can, in principle, create a phase-transition when certain conditions are met.
An important feature of tetrahedral liquids, as observed in simulations and experiments, is that the average coordination numbers changes significantly as a function of the temperature. For that reason we would like to propose an approximate statistical mechanics for the Stillinger-Weber liquid, based on a description of the local structures. This kind of theory was proposed a long time ago for liquids by Bernal and others \cite{64B,94SK} , and recently it was developed to describe the structural changes that occur in supercooled liquids \cite{procaccia}. This kind of ``upscaling'' is based on an assumption of the additivity of the energy of the local structure and on the fact that supercooled liquids are ergodic \cite{08EP}.

The Stillinger-Weber model \cite{84SW} is a widely used empirical model of silicon which assumes that the leading terms in the N-body quantum mechanical interaction are the two-body and three-body terms. The energy function is:
\begin{equation}
\mathcal{U} = \sum_{i<j}\nu_2(r_{ij}) + \sum_{i<j<k}\nu_3(r_{ij}),
\end{equation}
where $\nu_2$ is a Lennard-Jones-like potential with a finite cutoff, and $\nu_3$ is constructed such that it nullifies when the angles between the three neighbors are $\theta=109.47^o$. This property guarantees that the system has a diamond-cubic ground-state at atmospheric pressure. The exact details of the functions $\nu_2$ and $\nu_3$ can be found in ref. \cite{84SW}.
In order to take into account the contribution of the three body interactions we make the following approximation: we assume that the first shell of neighbors of each particle can be mapped to the first-shell of a particle in a BCC lattice oriented in an arbitrary direction. This is an approximation since in the liquid, particles can assume arbitrary positions. The BCC first-shell seems to be the best discretization of the liquid first-shell environment since the central particle has at most 8 neighbors.  This is consistent with simulations that use the SW potential, since it is possible to arrange $4$ particles around the central particle in a way that replicates the environment of a particle in a diamond-cubic lattice (the ground-state of the Stillinger-Weber potential). The values of the energies of the different states $i$ were calculated from the SW potential by placing particles at the different possible locations in the BCC first-shell. The energies were found to be:
\begin{eqnarray}
E_i =&& \{0,\ -\epsilon_0,\ -2\epsilon_0,\ -2\epsilon_0+\epsilon_1,\ -3\epsilon_0,\ -3\epsilon_0+2\epsilon_1,\nonumber\\
&& -4\epsilon_0, -4\epsilon_0+3\epsilon_1,\ -4\epsilon_0+4\epsilon_1,-5\epsilon_0+4\epsilon_1,\nonumber\\
&& -5\epsilon_0+6\epsilon_1,\ -6\epsilon_0+8\epsilon_1,\ -6\epsilon_0+9\epsilon_1,\nonumber\\
&& -7\epsilon_0+12\epsilon_1,\ -8\epsilon_0+16\epsilon_1\},
\label{energies}
\end{eqnarray}
\begin{figure}[h]
\includegraphics[width=0.30\textwidth]{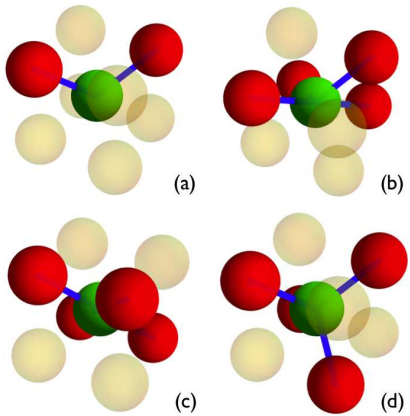}
\caption{Color online: representative nearest neighbor configuration for $n=2$ (a) and three possible configurations for $n=4$ (b)-(d), each having different three-body contributions to the potential energy. The central particle is in green, the neighbors are colored red and the rest of the possible particle positions are semi-transparent.}
\label{configurations}
\end{figure}
where $\epsilon_0$ is the typical energy scale of the two-body interaction, and $\epsilon_1$ is the energy scale of the three-body term. At this point $\epsilon_0$ and $\epsilon_1$ are left as free parameters. There are several different configurations which give the same energy and thus are degenerate. For a particle with no nearest neighbors there is only one possible configuration and thus the degeneracy of the state is $g_0=1$. When we have one nearest-neighbor in the first-shell, there are eight ways to arrange the neighbor in the BCC structure and $g_1=8$. When there are two nearest-neighbors there are twelve ways to arrange the particles in a configuration where the three-body term does not contribute to the energy, and sixteen different ways to arrange the two neighbors in a way were there is one contributions of the three-body term (the three-body term is non-zero only when there are at least three particles involved in the interaction). Similarly, we calculated the degeneracies that corresponds to the energies in Eq. (\ref{energies}):
\begin{equation}
g_i =\{ 1,8,12,16,8,48,2,32,36,8,48,12,16,8, 1\} \ .
\end{equation}
Each species occupies a typical volume in space, which depends on the geometry of the local configuration and the forces which the neighboring particles apply on the central particle. The typical volume per species was calculated by minimizing the energy for the different configurations and calculating the volume of the Voronoi cell of each particle using the {\it voro++} package given in \cite{07R}. The volume of each species increases with temperature (with a rate $\alpha$) due to thermal fluctuations while the three-body interactions cause the species to decrease its volume with temperature at a rate $\beta$ that depends on the number of three-body terms in the interaction. This is due to the repulsive nature of the three-body term which is more important at lower temperatures (fluctuations around the angle of minimum energy contribution are smaller). Both of these contributions were combined to give:
\begin{eqnarray}
v_i(T)=\{&1.395&  + \alpha\, T,1.297 + \alpha\, T,1.199 + \alpha\, T,\\ &1.220&  - \,\beta\,T + \alpha\, T, \nonumber \\ &1.101& + \alpha\, T,1.140- 2\,\beta\,T + \alpha\, T,1.0 + \alpha\, T,\nonumber\\&1.058& - 3\,\beta\,T + \alpha\, T,1.083 - 4\,\beta\, T + \alpha\, T,\nonumber\\&0.974& - 3\,\beta \,T + \alpha\, T,1.019 - 6\,\beta\, T + \alpha\, T,\nonumber\\&0.943& - 8\,\beta\, T + \alpha\, T,0.968 - 9\,\beta\, T + \alpha\, T ,\nonumber\\ &0.910& - 12\,\beta\, T  + \alpha\, T,0.870 - 16\,\beta\, T + \alpha\, T\,\},\nonumber
\end{eqnarray}
where $\alpha$ and $\beta$ have units of [$K^{-1}$]. The volumes are normalized by the volume of a particle in the ground-state at zero pressure (four neighbors in the Diamond-Cubic lattice - no contribution due to three-body interactions)  and are dimensionless.
The parameters $\alpha$ and $\beta$ together with $\epsilon_0$ and $\epsilon_1$ complete the parametrization of the model.
The statistical ensemble of particles at constant volume and temperature is given by the Helmholtz free energy:
\begin{equation}
F = U - TS = \sum_{k=0}^{14} N_kE_k   + T\sum_{k=0}^{14}N_k\ln \frac{N_k}{N g_k},
\end{equation}
where $S = -\sum_{k=0}^{14}N_k\ln (N_k/N g_k)$ is the configurational entropy. The free energy will have a minimum for the equilibrium values of the concentrations of species $p_k=\frac{N_k}{N}$. In order to account for the isochoric condition (constant volume) we need to demand that the total volume equals V. We do that by adding a volume constraint together with a Lagrange multiplier $\lambda$:
\begin{eqnarray}
f = && \sum_{k=0}^{14} p_k E_k  + T\sum_{k=0}^{14}p_k\ln \frac{p_k}{g_k} + \lambda (\sum_{k=0}^{14} p_k v_k - v),\,\,\,\,\,\,\,\,
\end{eqnarray}
where $v = \frac{V}{N}$ is the average volume per particle and $f=F/N$ is the free-energy per particle.
After minimizing the free-energy under the constraint we get the modified Boltzmann distribution:
\begin{equation}
\label{boltzmann}
p_i(T)=\frac{1}{\mathcal{Z}}g_ie^{-\frac{ E_i +  \lambda  v_i}{T}}\ , \quad
\mathcal{Z} (T) = \sum_i g_i e^{-\frac{ E_i + \lambda v_i}{T}},
\end{equation}
where $\mathcal{Z} (T)$ is the partition function. To that we need to add the constraint:
\begin{equation}
 \sum_i p_i(T) v_i(T) = v\label{constraint}.
\end{equation}
Note that different states $i$ may have the same coordination number $j$. We define the concentration $c_j(T)$ of the species having
coordination number $j$ at temperature $T$.
Solving numerically Eq. (\ref{boltzmann}) under the constraint (\ref{constraint}) we can find the average concentrations of the different species as a function of the temperature, from which we can compute $c_j(T)$. To asses the quality of the theory we present results from the NVT molecular dynamics simulations which were averaged over fifty different ensembles each containing $4096$ particles, see Fig. \ref{conc}. The theoretical predictions of the model, are obtained by
fitting the four free parameters ($\epsilon_0$,\ $\epsilon_1$, $\alpha$ and $\beta$). The results of are shown in the inset of Fig. \ref{conc}. Obviously the theory provides a semi-quantitative description of the numerical data.
\begin{figure}[h]
\centering
\includegraphics[width=0.5\textwidth]{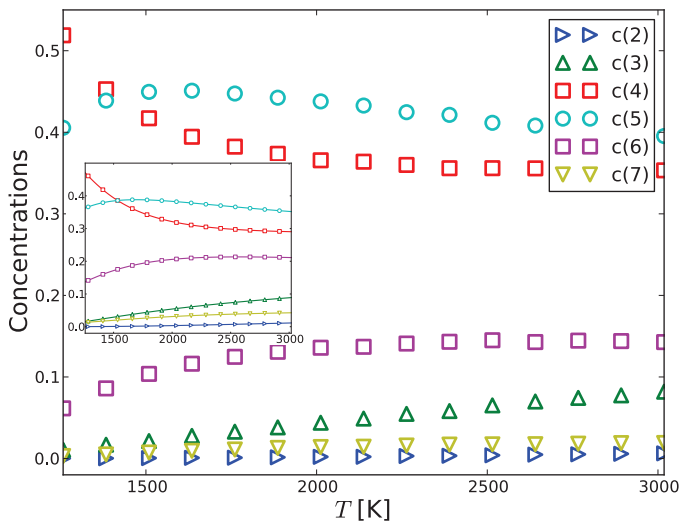}
\caption{Color online: Concentrations $c_j(T)$ of particles with coordination numbers $j$ from simulations. Inset: the theoretical results for comparison. The fitted values of the
parameters are $\epsilon_0=8.7\cdot10^{-6}$, $\epsilon_1=3.1\cdot10^{-6}$, $\alpha=1.4\cdot10^{-5}$ and $\beta=4.2\cdot10^{-6}$.}
\label{conc}
\end{figure}
Once we have fitted the free parameters to agree with the plots of the concentrations as a function of temperature, we do not
change them, and look at the model as fixed to assess its thermodynamic properties. As stated above, tetrahedral liquids exhibit peculiar low temperature behavior, such as a density maximum, an increase in the specific heat capacity and a negative thermal expansion. In figure \ref{simul}a we observe the calculated density $\rho$ for $P=-\frac{\partial F}{\partial V}=0$. A maximum appears at  about $T=950K$, roughly the same temperature as observed in \cite{05BM} for molecular dynamics of the SW model. We also observe a small density minimum at around $T=700K$ which is in agreement with  \cite{05BM} and with a recently published experiment on water \cite{07LZCMPC}. The thermal expansion and specific heat capacity at zero pressure are also shown in figures \ref{simul}b and \ref{simul}c. The specific heat shows a large increase while the thermal expansion decreases and becomes negative, as observed experimentally in water.
\begin{figure}[h]
\centering
\includegraphics[width=0.5\textwidth]{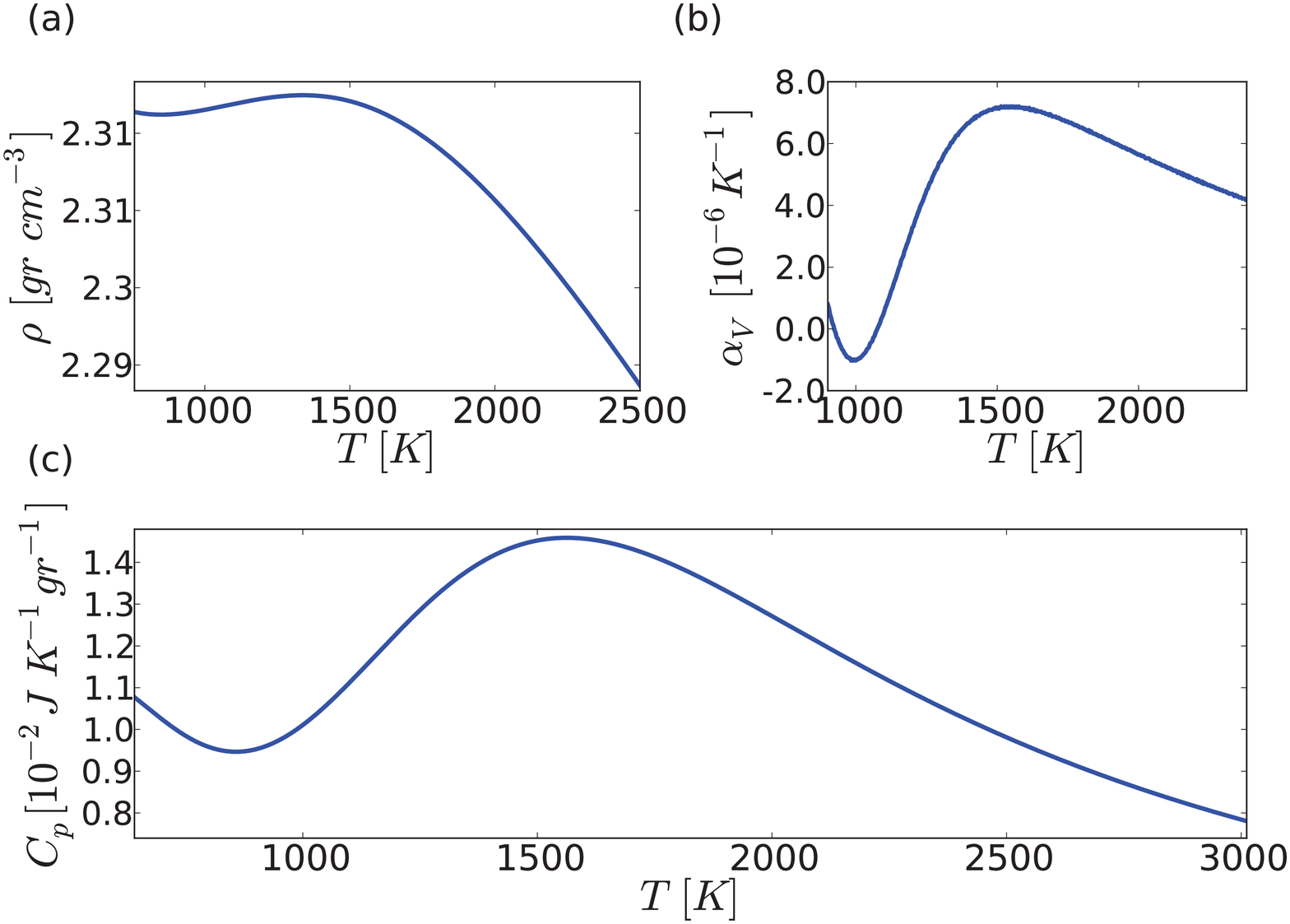}
\caption{Density maximum and minimum, thermal expansion minimum and specific heat maximum for $P=0$.}
\label{simul}
\end{figure}

\begin{figure}[h]
\centering
\includegraphics[width=0.5\textwidth]{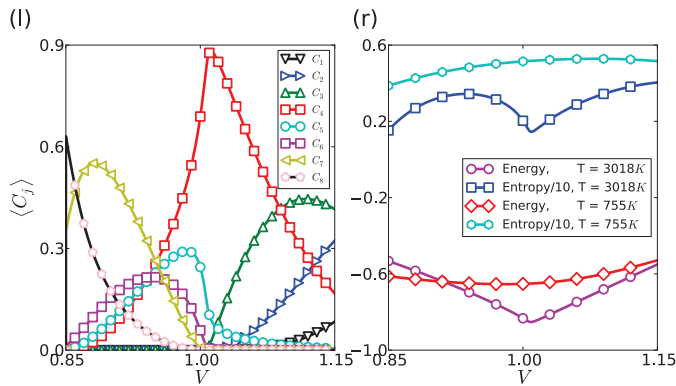}
\caption{Color online: (a) Concentrations of $C_j(T)$ as a function of volume for $T=755K$. (b) Comparison of the potential energy and the entropy for two different temperatures.}
\label{figure4}
\end{figure}

To understand these results, we consider the energy and the entropy as a function of the volume $V$ at two different temperatures ($T=755K$ and $T=3018K$). These can be immediately computed from the model, and provide us with insights into the physical mechanism that allows for the anomalous behavior shown in the previous subsection. Observe Fig. \ref{figure4}r: at high temperatures, the configurational entropy (divided by $10$ for comparison) has one maximum at a relatively high volume. The reason is that at a higher volume there are more ways to arrange species with different volumes. Since at high temperatures the term $-T\,S$ in the free energy is dominant, the minimum of the free-energy will appear at a higher volume. Lowering the temperature means that the potential energy becomes more important than the entropy and the equilibrium configurations tends to prefer the grounds-state species which have the lowest energy. Therefore, when the temperature is low the energy will prefer the volume V=1 and will develop a minimum around that point, as one can observe at Fig. \ref{figure4}r. When we constrain the ensemble to have a higher volume, more species appear in order to fill the gaps in space; similarly smaller species appear when the volume is smaller than unity(cf. Fig. \ref{figure4}l). The configurational entropy also develops a minimum around $V=1$ since most of the particles at this volume will only have four neighbors. Therefore, the minimum in the free-energy decreases and approaches the ground-state volume as the temperature is lowered. As we explained above, the increase in the concentrations of the various species when the volume is altered with respect to $V=1$, increases the configurational entropy which causes it to exhibit a double maximum, where a small increase in the volume results in a large increase in the entropy (cf. Fig. \ref{figure4}r). If this effect were sufficiently strong, the entropy could ``pull" the free-energy to higher volumes, despite the low temperature. {\it This causes the density to have a maximum} at a low temperature. If the increase in entropy due to volume changes were even larger, the free energy could develop a second minimum - a first order phase transition. In the present model the entropy contribution is not large enough to cause a phase transition. Beaucage at al. \cite{05BM} found that the liquid-liquid phase-transition in the Stillinger-Weber model is very sensitive to small perturbations of the potential and simulation parameters. When the three-body contribution increases by as little as $5\%$, the phase transition disappears while the system maintains the density-maximum property. Their result indicates that similar potentials can lead to a phase transition in some cases but not in others. A liquid can exhibit anomalous behavior without undergoing a phase transition. We suggest that this insight applies to real tetrahedral liquids: some may exhibit a phase transition while others may show anomalous behavior without a phase transition.

In summary, we have constructed a simple statistical mechanics theory which describes the local structure of a generic model of tetrahedral liquids, the Stillinger-Weber model. We showed that the theory, intended originally to describe only the structure, can account for the anomalous behavior of tetrahedral liquids and that this behavior can be understood as a competition between configurational entropy and potential energy. We suggest that the same physics can give rise to a liquid-liquid phase transition, and that it is possible that in some tetrahedral liquids the anomalous behavior is accompanied by a phase-transition while in others the phase transition does not occur. It is
important to stress however that the existence of a phase transition is {\it not} a pre-requisite for the thermodynamic anomalies which
are perfectly present even without any phase transition in the supercooled liquid.

This work has been supported in part by an ``ideas" advanced grant from the European Research Council, by the German Israeli Foundation and the Israel Science Foundation. I.R. was supported by the US Department of Energy through contract DE-AC52-06NA25396.

\end{document}